\documentclass[aps,prd,twocolumn,superscriptaddress,nofootinbib]{revtex4-2}

\usepackage[utf8]{inputenc}
\usepackage{amsmath}
\usepackage{amsfonts}
\usepackage{bm}
 \usepackage{booktabs}

\usepackage[separate-uncertainty=true,exponent-product=\cdot]{siunitx}
\usepackage{natbib}
\usepackage[bottom]{footmisc}
\usepackage[hidelinks]{hyperref}
\usepackage{mathtools}
\usepackage{listings}

\newcommand{\rmd}{{\rm d}}
\newcommand{\rmi}{{\rm i}}
\newcommand{\rme}{{\rm e}}
\newcommand\tighteq{\mkern1.5mu{=}\mkern1.5mu}
\newcommand{\mchirp}{{\mathcal{M}}}
\newcommand{\phiref}{\phi_{\rm ref}}
\newcommand{\dl}{D}
\newcommand{\tgeo}{t_\oplus}
\newcommand{\los}{\bm{\hat n}}
\newcommand{\intrinsic}{\bm \theta_{\rm int}}
\newcommand{\extrinsic}{\bm \theta_{\rm ext}}
\newcommand*{\timeseries}[1][t]{z_{mpd}\left(#1; \intrinsic\right)}
\newcommand{\covariance}{c_{mm'pp'd}(\intrinsic)}
\newcommand*{\tdet}[1][(\tgeo, \los)]{t_d#1}

\usepackage{xcolor}

\definecolor{codegreen}{rgb}{0,0.6,0}
\definecolor{codegray}{rgb}{0.5,0.5,0.5}
\definecolor{codepurple}{rgb}{0.58,0,0.82}
\definecolor{backcolour}{rgb}{0.95,0.95,0.92}
\lstdefinestyle{mystyle}{
    backgroundcolor=\color{backcolour},   
    commentstyle=\color{codegreen},
    keywordstyle=\color{magenta},
    numberstyle=\tiny\color{codegray},
    stringstyle=\color{codepurple},
    basicstyle=\ttfamily\footnotesize,
    breakatwhitespace=false,
    breaklines=true,
    captionpos=b,
    keepspaces=true,
    numbers=left,
    numbersep=5pt,
    showspaces=false,                
    showstringspaces=false,
    showtabs=false,
    tabsize=2
}
\begin{document}

\title{Fast marginalization algorithm for optimizing gravitational wave \texorpdfstring{\\}{}detection, parameter estimation and sky localization}

\author{Javier Roulet}
\affiliation{TAPIR, Walter Burke Institute for Theoretical Physics, California Institute of Technology, Pasadena, CA 91125, USA}
\email{jroulet@caltech.edu}

\author{Jonathan Mushkin}
\affiliation{\mbox{Department of Particle Physics \& Astrophysics, Weizmann Institute of Science, Rehovot 76100, Israel}}

\author{Digvijay Wadekar}
\affiliation{\mbox{School of Natural Sciences, Institute for Advanced Study, 1 Einstein Drive, Princeton, NJ 08540, USA}}

\author{Tejaswi~Venumadhav}
\affiliation{\mbox{Department of Physics, University of California at Santa Barbara, Santa Barbara, CA 93106, USA}}
\affiliation{\mbox{International Centre for Theoretical Sciences, Tata Institute of Fundamental Research, Bangalore 560089, India}}

\author{Barak Zackay}
\affiliation{\mbox{Department of Particle Physics \& Astrophysics, Weizmann Institute of Science, Rehovot 76100, Israel}}

\author{Matias Zaldarriaga}
\affiliation{\mbox{School of Natural Sciences, Institute for Advanced Study, 1 Einstein Drive, Princeton, NJ 08540, USA}}
             
\begin{abstract}
We introduce an algorithm to marginalize the likelihood for a gravitational wave signal from a quasi-circular binary merger over its extrinsic parameters, accounting for the effects of higher harmonics and spin-induced precession.
The algorithm takes as input the matched-filtering time series of individual waveform harmonics against the data in all operational detectors, and the covariances of the harmonics.
The outputs are the Gaussian likelihood marginalized over extrinsic parameters describing the merger time, location and orientation, along with samples from the conditional posterior of these parameters.
Our algorithm exploits the waveform's known analytical dependence on extrinsic parameters to efficiently marginalize over them using a single waveform evaluation.
Our current implementation achieves a 10\% precision on the marginalized likelihood within  $\approx\SI{50}{\milli\second}$ on a single CPU core and is publicly available through the package \href{https://github.com/jroulet/cogwheel}{\texttt{cogwheel}}.
We discuss applications of this tool for (\textit{i}) gravitational wave searches involving higher modes or precession, (\textit{ii}) efficient and robust parameter estimation, and (\textit{iii}) generation of sky localization maps in low latency for electromagnetic followup of gravitational-wave alerts. 
The inclusion of higher modes can improve the distance measurement, providing an advantage over existing low-latency localization methods.

\end{abstract}

\maketitle

\section{Introduction}
\label{sec:introduction}

Gravitational wave astronomy has undergone tremendous progress over recent years, made possible by the advent of the advanced LIGO \cite{Abbott2009} and Virgo \cite{Aasi2012} detectors.
In order to maximize the scientific impact of these extraordinary data, the community has developed advanced methods for identifying signals \cite{Klimenko2016, Usman2016, Messick2017, Nitz2017, Sachdev2019, Venumadhav2019, Cannon2021, DalCanton2021, Zackay2021, Ewing2024}, which have yielded over a hundred detections to date \cite{Abbott2019_GWTC1,Abbott2021_GWTC2,Abbott2023_GWTC3,Venumadhav2020,Olsen2022,Mehta2023,Wadekar2023,Nitz2019,Nitz2020,Nitz2021}; for estimating their source parameters \cite{Veitch2015,Cornish2015,Lange2018,Ashton2019,Ashton2021,Biwer2019,Chua2020,Breschi2021,Cornish2021,Gabbard2021, Dax2021,Roulet2022,Fairhurst2023,Tiwari2023,Wong2023}, which have provided invaluable insights into the astrophysics of compact binaries \cite{Abbott2023_GWTC3pop}; and for searching for short-lived electromagnetic counterparts \cite{Singer2016,Cornish2016,Sachdev2020,Magee2021,Pathak2024}, that enabled the identification of the kilonova from the binary neutron star merger GW170817 \cite{Abbott2017,Abbott2017b}.
One technique that has recurrently found applications in all these fronts is the ability to marginalize the signal's likelihood over a subset of its parameters.
In particular, extrinsic parameters describing the location of the observer relative to the source are the most amenable to marginalization, as their effect on the signal can be modeled analytically \cite{Pankow2015, Singer2016}.

In this study, we present an algorithm for marginalizing the likelihood for a gravitational wave signal from a quasi-circular binary merger over extrinsic parameters, assuming Gaussian noise and accounting for higher harmonics and spin-induced precession.
The inputs to the algorithm are a set of matched-filtering time series of the waveform against the data (one time series for each harmonic mode, polarization $\{+, \times\}$ and detector) and the covariances of these components.
The output is the Gaussian likelihood ratio marginalized over extrinsic parameters
\begin{equation}
    \label{eq:marg_like}
    \overline{\mathcal L}(\intrinsic)
        = \int \rmd \extrinsic \, \pi(\extrinsic) \mathcal L (\intrinsic, \extrinsic),
\end{equation}
with
\begin{equation}\label{eq:Gaussian_like_ratio}
    \mathcal{L(\bm \theta)}
    = \frac{p(d \mid \bm \theta, {\rm signal})}{p(d \mid \text{Gaussian noise})},
\end{equation}
where $d$ is the data, $\bm\theta \equiv (\intrinsic, \extrinsic)$ are the intrinsic and extrinsic parameters of the signal, and $\pi(\bm\theta) \equiv p(\bm \theta \mid \text{signal})$ is the prior distribution.
Since the analytical dependence on $\extrinsic$ is known, $\overline{\mathcal L}(\intrinsic)$ can be evaluated using a single waveform query.

We envision at least three applications of this algorithm: as a piece of the detection statistic in a search incorporating higher modes or precession, as a tool for efficient and robust parameter estimation, and as a means of producing sky localization maps in low latency for electromagnetic follow-up of gravitational-wave alerts.

In a search, according to the Neyman--Pearson lemma, the optimal detection statistic is the likelihood ratio $\Lambda$ between the two competing hypotheses, namely that there is a signal versus only noise:
\begin{equation}
    \Lambda(d) = \frac{p(d \mid {\rm signal})}{p(d \mid {\rm noise})}
    = \frac{\int \rmd \bm\theta \, p(d \mid \bm\theta, {\rm signal}) \, p(\bm\theta \mid {\rm signal})}{p(d \mid {\rm noise})}.
\end{equation}

By Eq.~\eqref{eq:Gaussian_like_ratio},
\begin{equation}
    \Lambda(d)
    = \int \pi(\bm\theta) \mathcal{L}(\bm\theta) \,\rmd \bm\theta \cdot\frac{p(d \mid \text{Gaussian noise})}{p(d \mid {\rm noise})}.
\end{equation}
In this work we study the marginalization of the likelihood over extrinsic parameters, under the assumption of Gaussian noise in order to outline a tractable, well-defined problem.
Later stages in the pipeline undertake the remaining marginalization over intrinsic parameters and apply a correction for the fact that, empirically, the noise distribution $p(d \mid \text{noise})$ is not Gaussian \cite{Wadekar2024}.
Most search pipelines have implemented techniques to perform the extrinsic-parameter marginalization, but they have generally assumed quadrupolar gravitational radiation and non-precessing sources \cite{Klimenko2016,Dhurandhar2017,Nitz2017,Hanna2020,Olsen2022}.
The contribution of this work is to include higher-order modes and spin-induced precession in the signal model, while maintaining a low computational cost compared to other components of the search.
Indeed, this algorithm has been crucial in a recent search including higher modes \cite{Wadekar2023}.

In the context of parameter estimation, in the traditional likelihood-based paradigm a stochastic sampler is used to explore the high-dimensional parameter space, by alternatingly proposing evaluation points and computing the posterior probability density.
Marginalizing the likelihood removes the extrinsic parameters from the problem, simplifying the task for the sampler. In particular, the extrinsic parameters tend to exhibit multiple modes and nonlinear degeneracies \cite{Singer2014, Roulet2022}.
This approach has been pursued in the literature \cite{Pankow2015}, but with implementations that either did not support higher modes and precession \cite{Islam2022}, or that were significantly more computationally intensive than the one we present here \cite{Wysocki2019}.

Finally, a byproduct of this algorithm is a set of extrinsic parameter samples weighted according to their conditional posterior probability, conditioned on the intrinsic parameters.
If one has estimates of the intrinsic parameters (e.g.\ from a search pipeline), this method can be used to measure the extrinsic parameters within seconds.
This mode of operation is similar to the \texttt{BAYESTAR} pipeline \cite{Singer2016}, except generalized to include precession and higher-order modes.
This is important because higher modes are sensitive to the inclination of the binary, potentially breaking its degeneracy with the distance and localizing the source to a smaller volume \cite{Abbott2020_GW190412}.
Higher modes may also improve the constraints on the mass ratio of the merging objects, informing about their nature and probability of sourcing an electromagnetic counterpart.

The article is organized as follows.
\S\ref{sec:method} provides a detailed description of the marginalization algorithm.
\S\ref{sec:convergence} studies its convergence and computational cost.
\S\ref{sec:applications} explores the applications to search, parameter estimation and low-latency source localization.
We conclude in \S\ref{sec:conclusions}.
Appendix~\ref{app:tricks} describes various computational optimizations.
Appendix~\ref{app:code} includes a code snippet demonstrating how to use our algorithm for parameter inference with the \texttt{cogwheel} software.\footnote{\label{footnote:cogwheel_url}\url{https://github.com/jroulet/cogwheel}}

\section{Method}
\label{sec:method}

\subsection{Summary of the algorithm for extrinsic-parameter marginalization}

Given the data and a choice of intrinsic parameters $\intrinsic$, we compute the marginalized likelihood Eq.~\eqref{eq:marg_like} using a combination of integration methods: we integrate over distance by interpolating a precomputed table, over orbital phase by trapezoid quadrature, and over the remaining extrinsic parameters using adaptive importance sampling.

We first generate a large number of samples for extrinsic parameters excluding distance and orbital phase (namely: sky location, geocenter time of arrival and polarization\footnote{And inclination, if one restricts to aligned spins.}). We draw these from a proposal distribution (described in \S\ref{sec:proposal}) designed to be easy to compute and sample from, and to approximately match the posterior conditional on the intrinsic parameters.
For each of these samples, we compute the complex inner products $( d \mid h^0_m )$ and $( h^0_m \mid h^0_{m'} )$ for a signal $h^0$ at a fiducial distance and orbital phase, where $h^0_{m}$ is the inertial-frame waveform that is generated by spherical harmonic modes with azimuthal index $m$ in the co-precessing frame.
These quantities transform in simple ways under a change of orbital phase or distance.
We use the trapezoid quadrature rule to integrate over phase, and a lookup table to integrate over distance.
Finally, we reweight each sample by the ratio of its posterior (marginalized over phase and distance) to the proposal probability.
This yields two useful products: an estimate of the likelihood marginalized over all extrinsic parameters, and a set of weighted samples from the extrinsic parameter posterior. If the proposal distribution is found to inadequately describe the posterior (diagnosed as a low effective sample size) we adaptively tune the proposal and produce additional samples until we achieve convergence.

\subsection{Waveform decomposition}

In this section we write the explicit dependence of the likelihood on extrinsic parameters. We find it convenient to express the model in terms of products of various tensors, each depending on a reduced set of parameters. Throughout, we will use the subindex $d$ to label detectors, $p$ for polarizations $\{+, \times\}$, and $(\ell, m)$ for co-precessing frame harmonic modes.
We will use the inner product between two time series defined as \cite{Finn1992}
\begin{equation}
    \langle x \mid y \rangle
    = 4 \Re \int_0^\infty \rmd f
        \,\frac{\tilde x(f) \,\tilde y^\ast(f)}{S(f)},
\end{equation}
where $S$ is the one-sided noise power spectrum, and a summation over detectors is assumed.

We start with the standard Gaussian likelihood ratio $\mathcal L$ (we henceforth refer to $\mathcal L$ simply as the likelihood):
\begin{equation}
    \label{eq:gaussian_like}
    \begin{split}    
        \ln \mathcal{L}(\bm\theta)
        &= \langle d \mid h(\bm\theta)\rangle
            - \frac 12 \langle h(\bm\theta) \mid h(\bm\theta) \rangle.
    \end{split}
\end{equation}
Here, $d$ is the strain data and $h$ the model waveform. 
Extrinsic parameters modify the waveform in a well-understood way:\footnote{We follow the default \texttt{LALSimulation} convention and use labels $m>0$, understanding that the $m$ and $-m$ coprecessing harmonics are summed together using $\tilde h_{\ell m}(f) = (-1)^\ell \tilde h_{\ell,-m}(-f)$, with $f > 0$ \cite{LALSuite,Kidder2008,Pratten2021}.}
\begin{multline}
\label{eq:waveform}
    \tilde h_{d}(f; \intrinsic, \psi, \los, \tgeo, \phiref, \dl)
    = \\\sum_{m=1}^{\ell_{\rm max}}
      \sum_{p \in \{+, \times\}}
        \tilde h_{mp}(f;\intrinsic)
        \frac{F_{dp}(\los, \psi)
        \rme^{- \rmi 2\pi f \tdet}
        \rme^{\rmi m \phiref}
        }{\dl}
\end{multline}
where we have grouped the harmonics by $m$:
\begin{equation}
    \tilde h_{mp}(f; \intrinsic)
    \coloneqq \sum_\ell \tilde h_{\ell m p}(f; \intrinsic, \dl \tighteq 1, \phiref \tighteq 0).
\end{equation}
Here, the indices $\ell, m$ denote spherical-harmonic modes in the co-precessing frame, but the harmonic $\tilde{h}_{\ell mp}$ is the inertial-frame waveform generated by the ``twisting up” procedure operating on this co-precessing harmonic \cite{Schmidt2012}. In particular, the $\tilde{h}_{\ell mp}$ has different spherical harmonic content in the inertial frame.

Reading Eq.~\eqref{eq:waveform} from the left, the right-hand side is interpreted as follows: the source emits polarized waves $h_{mp}$, to which the detector has an antenna response $F_{dp}(\los, \psi)$ that depends on the geometrical configuration.
The signal arrives at each detector at time
\begin{equation}\label{eq:tdet}
    \tdet = \tgeo - \bm r_d \cdot \los / c,
\end{equation}
which is the overall time of arrival at geocenter $\tgeo$ plus a time-of-travel correction that depends on the location $\bm r_d$ of the detector projected onto the line of sight $\los$.
Each co-precessing harmonic $\tilde h_{\ell m}$ transforms according to $\rme^{\rmi m \phiref}$ under a rotation in the plane of the binary.\footnote{When we vary the orbital phase $\phiref$ we hold the black hole spins fixed with respect to the orbital angular momentum and the direction of propagation (not with respect to the orbital separation vector) \cite{Roulet2022}.}
Lastly, the waveform amplitude decays in inverse proportion to the luminosity distance to the source, $D$.

For precessing signals, the inclination of the orbit is frequency-dependent, and therefore we will treat it as an intrinsic (non-marginalized) parameter.
For non-precessing (aligned-spin) systems, the inclination can be treated analytically by replacing $\rme^{\rmi m \phiref}$ by the spin-weighted harmonic $_{-2}Y_{\ell m}(\iota, \phiref)$ in Eq.~\eqref{eq:waveform} \cite{Thorne1980}.

Using Eq.~\eqref{eq:waveform}, we can rewrite Eq.~\eqref{eq:gaussian_like} in terms of factors that depend separately on the intrinsic or the extrinsic parameters.
\begin{multline}\label{eq:dh}
    \langle d \mid h \rangle 
    = \\\frac 1D
    \Re\bigg\{
    \sum_m \rme^{-\rmi m \phiref} 
    \sum_{d,p}
    F_{dp}(\los, \psi)
    \timeseries[\tdet]
    \bigg\},
\end{multline}
where the time series
\begin{equation}
    \label{eq:timeseries}
    \timeseries
    \coloneqq 4 \int_0^\infty \rmd f
        \frac{\tilde d_d(f) \tilde h^\ast_{mp}(f; \intrinsic)}{S_d(f)}
        \rme^{\rmi 2 \pi f t}
\end{equation}
is the complex matched-filter output \cite{Allen2012} of the waveform's mode $m$ and polarization $p$ in the $d$th detector. In practice only a short interval of time around the peak is needed. Similarly,
\begin{equation}\label{eq:hh}
    \begin{split}
        \langle h \mid h \rangle
        &= \frac{1}{D^2}
            \sum_{m,m'} \rme^{\rmi (m' - m) \phiref}\\
        &\quad\times \sum_{d,p,p'}
            \covariance F_{dp}(\los, \psi) F_{dp'}(\los, \psi)
    \end{split}
\end{equation}
with
\begin{equation}
    \label{eq:covariance}
    \covariance
    = 4 \int_0^\infty \rmd f
        \frac{\tilde h_{mp}(f; \intrinsic)
              \tilde h^\ast_{m'p'}(f; \intrinsic)}
              {S_d(f)}.
\end{equation}
Substituting Eqs.~\eqref{eq:dh} and \eqref{eq:hh} into \eqref{eq:gaussian_like}, we have decomposed the likelihood into factors that depend either on extrinsic or intrinsic parameters.
The matched-filter time series $\timeseries$ and covariances $\covariance$ encapsulate all the dependence on intrinsic parameters and will be the inputs to our computation.

\subsection{Phase and distance integration}
\label{sec:phase}

The distance, orbital phase and polarization are simpler to marginalize than other extrinsic parameters, because they do not affect the times of arrival, and hence their effect on the waveform is independent of frequency.
This makes it inexpensive to vary these parameters, as the data can first be compressed to a few numbers with the frequency axis collapsed.
Here we will marginalize the distance and orbital phase explicitly, and in \S\ref{sec:proposal} we will integrate the remaining extrinsic parameters using importance sampling.
Since the orbital phase and polarization are largely degenerate \cite{Veitch2015, Roulet2022}, it suffices to marginalize only one of the two at high resolution; we will do the phase.

Holding all other parameters fixed, the distance-marginalized likelihood $\overline{\mathcal{L}}_D$ can be expressed in terms of only two values, namely the inner products $\langle d \mid h_1 \rangle$ and $\langle h_1 \mid h_1 \rangle$ for a waveform at unit distance, $h_1 \coloneqq h(D{=}1)$:
\begin{multline}
\label{eq:marg_dist}
    \overline{\mathcal{L}}_D\big(\langle d \mid h_1 \rangle,
                             \langle h_1 \mid h_1 \rangle\big)\\
    = \int \rmd D \, \pi(D) \exp\left(
        \frac{\langle d \mid h_1 \rangle}{D}
        -\frac{\langle h_1 \mid h_1 \rangle}{2D^2}\right).
\end{multline}
Following \citet{Singer2016}, after suitable rescaling and reparametrization Eq.~\eqref{eq:marg_dist} can be efficiently evaluated by 2D interpolation of a precomputed lookup table.
This is possible since neither higher modes nor precession modify the dependence of the waveform on distance.

However, higher modes do change the dependence on orbital phase, and hence we cannot marginalize the phase analytically, as usually done for quadrupolar waveforms.
Instead, we use trapezoid quadrature, which performs adequately since the likelihood is a periodic function of the orbital phase.
To integrate Eq.~\eqref{eq:marg_dist}, it suffices to evaluate $\langle d \mid h_1 \rangle$ and $\langle h_1 \mid h_1 \rangle$ on a regular grid $\{\phi_{{\rm ref}, o}\}$ covering the orbital phases, where the  subindex $o$ runs through $1,\ldots, N_{\phi}$.
For $\langle d \mid h_1 \rangle$, we compute
\begin{equation}\label{eq:dh_io}
    \langle d \mid h_1 \rangle_{o}
        = \Re\sum_m ( d \mid h_1 )_{m} \Phi_{mo},
\end{equation}
where
\begin{equation}
    ( d \mid h_1 )_{m}
        \coloneqq \sum_{p,d} F_{dp}(\los, \psi) \timeseries[\tdet[(\tgeo, \los)]]
\end{equation}
is obtained by cubic spline interpolation of the time series $z_{mpd}$ and 
\begin{equation}
    \Phi_{mo} \coloneqq \exp(\rmi \,m\, \phi_{{\rm ref}, o})
\end{equation}
is precomputed.
For $\langle h_1 \mid h_1 \rangle$, similarly
\begin{align}
    \langle h_1 \mid h_1 \rangle_{o}
        &= \Re\sum_{m, m'} ( h_1 \mid h_1 )_{mm'} \Phi_{mm'o} \label{eq:hh_io},\\
    ( h_1 \mid h_1 )_{mm'}
        &\coloneqq \sum_{d,p,p'} \covariance F_{dp}(\los, \psi) F_{dp'}(\los, \psi)\\
    \Phi_{mm'o} &\coloneqq \exp\big[\rmi (m' - m) \phi_{{\rm ref}, o}\big].
\end{align}
Using Eqs.~\eqref{eq:marg_dist}, \eqref{eq:dh_io} and \eqref{eq:hh_io} we obtain the likelihood marginalized over orbital phase and distance:
\begin{equation}
    \label{eq:phase_quadrature}
    \overline{\mathcal{L}}_{\phi D}(\psi, \los, \tgeo; \intrinsic)
    \approx \frac{1}{N_\phi}\sum_{o=1}^{N_\phi}
        \overline{\mathcal{L}}_D\big(
            \langle d \mid h \rangle_{o},
            \langle h \mid h \rangle_{o}\big).
\end{equation}
From these data products, posterior samples of distance and orbital phase can also be readily generated.
Phase samples can be drawn from the grid according to the weights (the summands in Eq.~\eqref{eq:phase_quadrature}); and then distance samples can be produced from the integrand of Eq.~\eqref{eq:marg_dist} with inverse transform sampling.

\subsection{Time, sky location and polarization integration}
\label{sec:proposal}

We perform the integral in Eq.~\eqref{eq:marg_like} over the remaining extrinsic parameters (sky location $\los$, time of arrival $\tgeo$ and polarization angle $\psi$) using importance sampling \cite{Owen2013, Roulet2024}.
We will choose a proposal distribution $p(\psi, \tgeo, \los)$, and construct it in a way that will allow us to easily generate samples from it.
The marginal likelihood will be estimated from those samples as
\begin{align}
\label{eq:importance_marg_like_0}
    \overline{\mathcal{L}}(\intrinsic)
    &\approx
    \frac1N \sum_{i=1}^N
    \frac{\pi(\psi^i, \tgeo^i, \los^i)
          \,\overline{\mathcal{L}}_{\phi D}(\psi^i, \los^i, \tgeo^i; \intrinsic)}
         {p(\psi^i, \tgeo^i, \los^i)}.
\end{align}
The weighted samples also allow to sample the conditional posterior $p(\extrinsic \mid d, \intrinsic)$, by simply drawing according to the weights (i.e., the summands in Eq.~\eqref{eq:importance_marg_like_0}).

The variance of the estimator in Eq.~\eqref{eq:importance_marg_like_0} is highly sensitive to the choice of proposal: it vanishes when $p$ is proportional to the integrand (i.e., the conditional posterior for $(\psi, \tgeo, \los)$ in the numerator of Eq.~\eqref{eq:importance_marg_like_0}), on the other hand, it diverges if $p$ has a tighter support.
When this happens, a small number of samples in the tail of $p$ are disproportionately upweighted and dominate the sum.
Thus, we will design the proposal to approximately match the conditional posterior, erring on the side of having heavier tails.
Having reduced the dimensionality of the quasi-Monte Carlo integral Eq.~\eqref{eq:importance_marg_like_2} by explicitly integrating out the orbital phase and distance (\S\ref{sec:phase}) improves its efficiency, especially considering that the orbital phase is very well measured when other parameters are kept fixed.

For $\psi$, we simply use its uniform prior as proposal, based on the heuristic that it is rarely well constrained due to degeneracy with the orbital phase $\phiref$ \cite{Veitch2015,Roulet2022}.

In contrast, $\los$ and $\tgeo$ are usually measured very well compared to the size of their prior, calling for a more sophisticated proposal.
The constraints on these parameters are largely driven by the  measurement of the arrival times at the individual detectors.
Notably, it is possible to estimate these arrival times separately at each detector, and, furthermore, to precompute their relation to $\tgeo$ and $\los$ independently of the data.
With these insights, we follow \cite{Olsen2022, Islam2022} and specify our proposal distribution over $(\tgeo, \los)$ with the help of an auxiliary proposal $P(\bm\tau)$ for the discretized times of arrival $\tau_d$ at each detector.\footnote{We use capital letters for discrete distributions, lowercase for continuous distributions.}

\subsubsection{Proposal for arrival direction and geocenter time}
\label{sec:tgeo_los_proposal}

We choose a timescale $\Delta$ sufficiently small to resolve the autocorrelation length of the whitened template (thus, structure in the matched-filtering time series), and discretize the time axis at this resolution.
We then partition the $(\tgeo, \los)$ space into exhaustive disjoint regions $\mathcal D(\bm\tau)$, where $\bm\tau \equiv \{\tau_d\}$ defines a discrete time of arrival at each detector, and $\mathcal{D}(\bm\tau)$ is the domain of arrival time and sky location consistent with those $\bm\tau$.
Our criterion for consistency is that the time of arrival at the first detector, and the time delays between the first detector and the others, match those of $\bm \tau$ to a precision $\Delta/2$:
\begin{multline}
    \mathcal{D}(\bm \tau) = \\\bigg\{\tgeo, \los:
    \big|t_{d_0}(\tgeo, \los) - \tau_{d_0}\big| < \frac\Delta2
    \;\land\;
    \big|\delta t_d(\los) - \delta\tau_d \big| < \frac\Delta2
    \bigg\}, \label{eq:D}
\end{multline}
with
\begin{align}
    \delta t_d(\los) &\coloneqq \tdet - t_{d_0}(\tgeo, \los) \\
    \delta\tau_d &\coloneqq \tau_d - \tau_{d_0},
\end{align}
where $d_0$ is the arbitrary first detector.
The time delays $\bm {\delta t}$ and $\bm{\delta\tau}$ have $N_{\rm detectors} - 1$ components each, and $\bm {\delta t}$ is independent of $\tgeo$.

Our strategy is to first draw samples $\bm \tau^i$ from a proposal $P(\bm\tau)$ (described later), and to each assign a $\tgeo^i, \los^i \sim \pi(\tgeo, \los \mid \bm \tau^i)$ drawn from the restricted prior, by means of a precomputed mapping that we construct as follows.

\begin{figure}
    \centering
    \includegraphics[width=\linewidth]{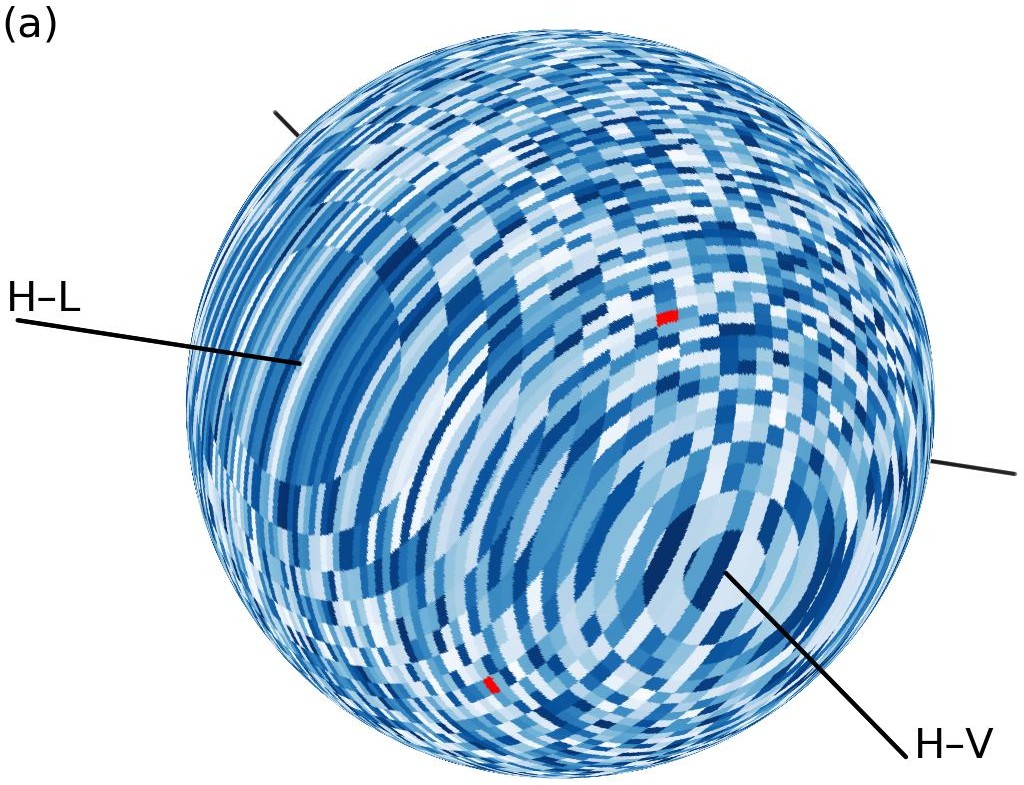}
    
    \vspace{10pt}
    \includegraphics[width=\linewidth]{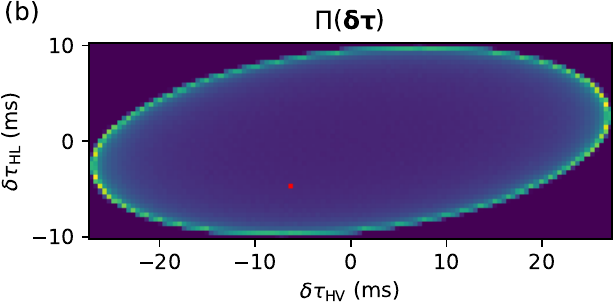}
    \caption[]{
    Partition of the space of arrival directions $\los$ by discretized time delays between detectors $\bm{\delta\tau}$, for the particular case of a Hanford--Livingston--Virgo network.
    We use this mapping to efficiently assign a consistent arrival direction to a proposed set of discrete arrival times at the detectors, during the importance-sampling marginalization over $\los$.
    (a) The time delay in each pair of detectors defines a ring in the sky, perpendicular to the corresponding detector separation vector (black axes).
    For the 3-detector network shown here, each subregion is the intersection of two rings.
    For a 2-detector network, the subregions would instead be annular, and for a single detector there would be one region covering the entire sky.\textsuperscript{\ref{footnote:multidet}}
    (b) Prior for discretized time delays between detectors, proportional to the solid angle of the associated patch of the sky.
    One particular $\bm{\delta\tau}$ is highlighted in red in both plots.
    Two sky patches, symmetric about the plane containing the detectors, share the same delays.
    The resolution of the map was lowered for visual clarity; by default we use a $4\times$ higher one of \SI{8192}{\hertz}.
    For a 2-detector network the prior would be a 1-dimensional array, and for a single detector it would be a scalar number.
    }
    \label{fig:skydict}
\end{figure}

Ahead of time, we draw a large number of samples (\num{e6} is our current default) isotropically distributed in the sky, in terms of Earth-fixed coordinates (latitude and longitude).
For each sample we compute $\bm{\delta t}(\los)$, and based on this we assign it to the nearest discretized time-delay $\bm{\delta \tau}$.
The resulting map is shown in Fig.~\ref{fig:skydict} for the example case of a Hanford--Livingston--Virgo network.
Given a $\bm{\delta\tau}$, the mapping provides a set of consistent sky location samples (color-coded in Fig.~\ref{fig:skydict}a).
Once a detector time sample $\bm{\tau}^i$ has been proposed, we compute $\bm{\delta \tau}^i$ and assign a sky location sample $\los^i$ from the corresponding entry of the mapping.
We also draw a time of arrival at the first detector $t_{d_0}^i$ uniformly within $\tau_{d_0}^i\pm \Delta/2$, and solve Eq.~\eqref{eq:tdet} to obtain $\tgeo^i$.

\footnotetext{For upcoming detector networks consisting of more than three detectors, the strategy of proposing arrival times at each detector would result in over-constrained sky locations.
The method can be generalized by using only the most sensitive three detectors for the proposal (and all the detectors for reweighting) in that case. \label{footnote:multidet}}

As a useful byproduct, this map allows us to obtain the physical prior on the time-delays $\Pi(\bm{\delta\tau})$, estimated as the fraction of samples assigned to each $\bm{\delta\tau}$.
We show this prior in Fig.~\ref{fig:skydict}b, it will be necessary later.
In particular, unphysical $\bm{\delta\tau}$ (say, delays longer than the gravitational-wave travel time between detectors) get no sky location samples assigned.
To keep the variance of this estimate low, we use a quasirandom Halton sequence to draw the sky locations, which covers the sky more uniformly than e.g.\ random sampling or a spherical grid.

Note that near the plane containing the detectors, perpendicular displacements produce only quadratic shifts in the arrival times.
Accordingly, in Fig.~\ref{fig:skydict}a these cells are elongated and cover a large solid angle, and have a large prior $\Pi(\bm{\delta\tau})$ in Fig.~\ref{fig:skydict}b.
We expect that sources near the plane of the detectors will have a relatively poor sky location measurement perpendicular to the plane.

The proposal density that results from the above process is
\begin{equation}\label{eq:tgeo_los_proposal}
    \begin{split}
        p(\tgeo^i, \los^i)
        &= \sum_{\bm\tau} P(\bm \tau) \pi(\tgeo^i, \los^i \mid \bm \tau) \\
        &= P(\bm \tau^i) \pi(\tgeo^i, \los^i \mid \bm \tau^i).
    \end{split}
\end{equation}
The second line follows because the restricted prior is zero for detector arrival times that are inconsistent with the sample: $\pi(\tgeo^i, \los^i \mid \bm\tau {\neq} \bm \tau^i) = 0$.
Combining Eqs.~\eqref{eq:importance_marg_like_0} and \eqref{eq:tgeo_los_proposal}, we obtain
\begin{equation}
\label{eq:importance_marg_like_preliminary}
    \overline{\mathcal{L}}(\intrinsic)
    \approx
    \frac1N \sum_{i=1}^N
    \frac{\pi(\tgeo^i, \los^i)
          \,\overline{\mathcal{L}}_{\phi D}(\psi^i, \los^i, \tgeo^i; \intrinsic)}
         {P(\bm\tau^i)\, \pi(\tgeo^i, \los^i \mid \bm \tau^i)}.
\end{equation}

The prior in the numerator of Eq.~\eqref{eq:importance_marg_like_preliminary} is separable and uniform, $\pi(\tgeo)\pi(\los) = \rm const$. Since there is no natural domain for the time, we will adopt a dimensionless prior $\pi(\tgeo) = 1$ and recognize that the marginalized likelihood has units of time.
The restricted prior $\pi(\tgeo, \los \mid \bm \tau)$ is proportional to $\pi(\tgeo, \los)$ but integrates to 1 over $\mathcal{D}(\bm \tau)$, hence, their ratio in Eq.~\eqref{eq:importance_marg_like_preliminary} is
\begin{equation}\label{eq:D_vol}
    \begin{split}
        \frac{\pi(\tgeo, \los)}{\pi(\tgeo, \los \mid \bm \tau)}
        &\equiv \Pi(\bm{\tau}) \\
        &= \int_{\mathcal{D}(\bm \tau)}\rmd \tgeo\rmd \los  \, \pi(\tgeo) \pi(\los) \\
        &= \Delta \cdot \Pi(\bm{\delta \tau}).
    \end{split}
\end{equation}

In Eq.~\eqref{eq:D_vol}, the $\tgeo$ integral equals $\Delta$, and the $\los$ integral yields $\Pi(\bm{\delta\tau})$, i.e.\ the fraction of the sky compatible with the time delays that we introduced in Fig.~\ref{fig:skydict}b.

Subtituting Eq.~\eqref{eq:D_vol} in \eqref{eq:importance_marg_like_preliminary}, we arrive at
\begin{equation}\label{eq:importance_marg_like}
    \overline{\mathcal{L}}(\intrinsic)
    \approx
    \frac1N \sum_{i=1}^N
    \frac{\Delta \cdot \Pi(\bm{\delta\tau}^i)}{P(\bm\tau^i)}
         \;\overline{\mathcal{L}}_{\phi D}(\psi^i, \los^i, \tgeo^i; \intrinsic).
\end{equation}

\subsubsection{Adaptive multiple importance sampling}

As discussed above, the variance of the importance sampling integral can be large if the proposal is misspecified.
The most sensitive component of the proposal is the auxiliary distribution of discrete detector arrival times $P(\bm \tau)$, as it is responsible for the largest reduction in phase space volume.
To make $P(\bm \tau)$ robust to an eventual initial misestimation, we will allow the option of adapting it as needed by iteratively proposing distributions $P^{(j)}$, that attempt to cover any problematic regions where the previous proposals were too narrow.
The total proposal is a mixture of the form
\begin{equation}\label{eq:proposal}
    \begin{split}
        P(\bm \tau) &= \sum_j \alpha_j P^{(j)}(\bm \tau) \\
        &= \sum_j \frac{N_j}{N} \prod_d P^{(j)}_d(\tau_d),
    \end{split}
\end{equation}
with $\sum_\tau P^{(j)}_d(\tau) = 1$.
That is, we define a series of adaptive proposals $P^{(j)}(\bm\tau)$, each factorizable over detectors.\footnote{Note that the total proposal $P(\bm\tau)$ is not factorizable.}
This property makes drawing samples of $\bm \tau \sim P^{(j)}$ a simple task, as the $\tau_d$ are drawn independently from one-dimensional distributions.
This task can be achieved with the inverse transform sampling technique, which is efficient and furthermore facilitates the use of quasi-Monte Carlo integration, as we will explain in \S\ref{sec:qmc}.
Every time we add a new proposal $P^{(j)}$, we draw $N_j$ samples from it, and combine them with the previous ones using the so-called balance heuristic $\alpha_j = N_j/N$, with $N = \sum_j N_j$.
Altogether, Eqs.~\eqref{eq:importance_marg_like} and \eqref{eq:proposal} become
\begin{align}
    \overline{\mathcal{L}}(\intrinsic)
    &\approx
    \sum_{i=1}^N w_i, \label{eq:importance_marg_like_2}\\
    w_i &= 
    \frac{\Delta \cdot \Pi(\bm{\delta\tau}^i)}{\sum_j \big( N_j \prod_d P^{(j)}_d({\tau_d}^i) \big )}
    \overline{\mathcal{L}}_{\phi D}(\psi^i, \los^i, \tgeo^i; \intrinsic). \label{eq:wi}
\end{align}
The importance sampling weights $w_i$ can be used to estimate the effective number of samples
\begin{equation}\label{eq:neff}
    N_{\rm eff} \equiv \frac{\big(\sum_i w_i\big)^2}{\sum_i w_i^2}.
\end{equation}
We set a threshold $N_{\rm eff}^{\rm min}$, and iteratively add proposals $P^{(j)}$ until the effective sample size meets this threshold (or a maximum number of proposals is reached).
Every time a new $P^{(j)}$ is added, we draw $N_j$ samples from it, update the weights of all samples and recompute $N_{\rm eff}$.
In \S\ref{sec:convergence} we will confirm that $N_{\rm eff}$ is a good tracer of the precision of the importance sampling integral.

\subsubsection{Initial proposal distribution of arrival times at detectors}
\label{sec:initial_proposal}

We choose the initial ($j=0$) proposal distribution of detector arrival times for each detector $d$ based on the matched-filtering time series and covariances.
We adopt the following functional form:
\begin{equation}
    P^{(0)}_d(\tau) = \Pi_d(\tau) \exp\big[\beta_d \ln \hat{\mathcal L}_d(\tau)\big],
\end{equation}
where $\hat{\mathcal L}_d$ is an approximate likelihood, $0<\beta_d\leq1$ is a tempering factor and $\Pi_d$ is a prior. For the likelihood we use
\begin{equation}\label{eq:time_likelihood}
    \ln\hat{\mathcal L}_d(\tau)
    = \frac{\left(\sum_{m,p} \big|z_{mpd}(\tau)\big|\right)^2}
                   {2\sum_{m,p}c_{mmppd}}.
\end{equation}
This expression follows from approximating that different modes and polarizations are orthogonal and have independent phases, and then maximizing Eq.~\eqref{eq:gaussian_like} over these phases and the distance.

The tempering factors $\beta_d$ are intended to make the initial proposal broader, and therefore more robust against missing the support of the posterior.
How to choose them depends on the application, for example in a search they may be fixed heuristically (e.g.\ $\beta = 0.5$), or in parameter estimation they can be tuned at the beginning to maximize $N_{\rm eff}$ for a reference waveform.

Finally, we use the priors $\Pi_d$ to incorporate information about the physically allowed time delays between detectors:
if the signal is weak in a detector, the arrival time at that detector might be meaningfully constrained by data in other detectors.
While the true prior on arrival times $\tau_d$ is correlated among detectors, in order to draw arrival time samples easily we require that the proposal distributions $P_d^{(j)}$ are uncorrelated (see Eq.~\eqref{eq:proposal}).
We circumvent this by conditioning the proposal in one detector on the other detectors' proposal distributions rather than on the individual values of the arrival time samples.
We achieve this as follows.
Once we have computed the likelihood \eqref{eq:time_likelihood}, we sort the detectors by decreasing $\max_\tau \hat{\mathcal L}_d$.
In the first (loudest) detector $d=1$, where the likelihood best constrains the time of arrival, we use a uniform prior
\begin{equation}
\Pi_{1}(\tau) = \text{const},
\end{equation}
this defines $P^{(0)}_1(\tau)$.
For the second detector we condition the prior on our knowledge of $P^{(0)}_1$: 
\begin{equation}\label{eq:Pi2}
\Pi_2 = P^{(0)}_1 \ast \Pi_{21},
\end{equation}
i.e.\ we use the proposal for the time of arrival at the first detector convolved with the prior distribution of time delays $\tau_{21} = \tau_2-\tau_1$ to the second detector.
This incorporates the information that there is a maximum allowed time delay.
We compute the prior for the arrival time at the third detector in a conceptually similar way, where now $\Pi_3$ is informed by $P^{(0)}_1$ and $P^{(0)}_2$.
We first estimate the time delay between the first two detectors by cross-correlating their proposals: $P(\tau_{21}) = P^{(0)}_1 \star P^{(0)}_2$.
We marginalize over this distribution to obtain $\Pi(\tau_{31} \mid P(\tau_{21})) = \sum_{\tau_{21}}\Pi(\tau_{21}, \tau_{31})P(\tau_{21})$, and finally arrive at
\begin{equation}
    \Pi_3 = P_1^{(0)}\ast\Pi_{31\mid 21}.
\end{equation}

\subsubsection{Adaptation}
\label{sec:adaptation}

After iteration $J$ of the adaptation, we have the set of samples $\{\bm\tau^i\}, i=\{1, \ldots, \sum_j^J N_j\}$ proposed so far, along with their weights $w_i$.

For the following proposal $P_d^{(J+1)}(\tau_d)$ we aim to match the proposal to the posterior, perhaps with heavier tails.
We obtain a measurement of the detector arrival time posterior by kernel density estimation (KDE) on the existing samples: we construct a histogram of $\{\tau_d^i\}$ weighted by $w_i$ and convolve it with a suitably chosen kernel.
We use a heavy-tailed Cauchy kernel $K(\delta \tau; \Sigma) \propto (\delta \tau^2 + \Sigma^2)^{-1}$.
We set the kernel width $\Sigma$ in each detector using Silverman's rule of thumb \cite{Silverman1986} with a lower bound $\Delta$: $\Sigma_d = \max\big\{\Delta, \left(4N_{\rm eff}/3\right)^{-1/5}\sigma_d\big\}$,
where $\sigma_d$ is the weighted standard deviation of the sample of $\{\tau_d^i\}$.

As a new proposal we use a hybrid between this KDE and the previous proposal:
\begin{equation}
    P^{(J+1)}_d(\tau_d)
    = \frac12\bigg( {\rm KDE}(\tau_d) + \sum_j^JP_d^{(j)}(\tau_d) \bigg).
\end{equation}
This increases the stability of the adaptation and handles satisfactorily the generic situation in which the original proposal was adequate in some detectors and not others.

\begin{figure}
    \centering
    \includegraphics[width=\linewidth]{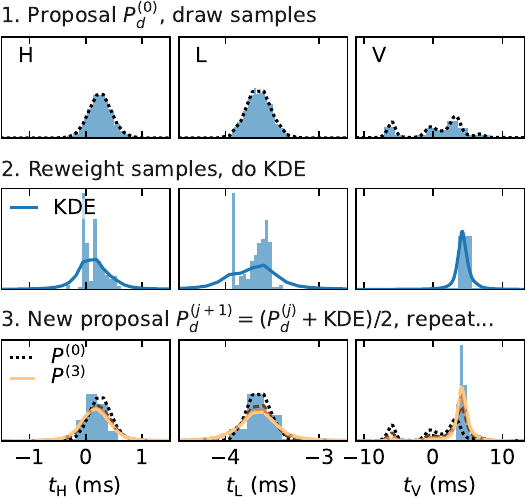}
    \caption{To reduce the variance of the importance sampling marginalization, we use an adaptive proposal distribution for the detector arrival times. \textit{Top:} An initial proposal is generated in each detector from the matched-filtering time series (\S\ref{sec:initial_proposal}).
    Sets of discrete arrival times $\bm\tau$ are sampled from these with quasi Monte Carlo.
    Physical parameters are assigned to each set of detector arrival times (\S\ref{sec:tgeo_los_proposal}).
    \textit{Center:} The physical samples are reweighted according to the ratio of their (coherent) posterior to the proposal (Eq.~\eqref{eq:wi}), and used to estimate the probability density in each detector via KDE (\S\ref{sec:adaptation}).
    \textit{Bottom:} The proposal is updated by averaging it with the KDE.
    Previous samples are kept, and their weights updated to reflect the change of the proposal.
    The process is repeated until the effective number of samples is satisfactory.
    }
    \label{fig:adaptation}
\end{figure}

This adaptation step is illustrated in the second and third rows of Fig.~\ref{fig:adaptation}.

\subsubsection{Quasi-Monte Carlo}
\label{sec:qmc}

In order to further reduce the variance of the $(\psi, \los,\tgeo)$ integral estimated in Eq.~\eqref{eq:importance_marg_like_2}, we jointly draw the samples of detector arrival times, polarization and subgrid time shift using quasi-Monte Carlo \cite{Owen2013}.
By design, the proposal is factorizable in all these variables. In each dimension we use inverse transform sampling, i.e., use the cumulative of the proposal, $u$, as coordinate, so that the proposal becomes the uniform distribution:
\begin{equation}\label{eq:u}
\begin{split}
    u_a(x_a) &\coloneqq \int_{-\infty}^{x_a} p_a(x'_a) \rmd x'_a, \\
    \Rightarrow u_a &\sim \rm U(0, 1),
\end{split}
\end{equation}
where $a$ labels each dimension (arrival time at each detector, subgrid timeshift, polarization) and $p_a$ is the corresponding proposal. Instead of drawing the $\{\bm u^i\}$ independently, we select them according to a scrambled Halton sequence. This introduces correlations between the samples, that decrease the variance of the estimator Eq.~\eqref{eq:importance_marg_like_2} by making the average covariance of the weights negative.
We then invert Eq.~\eqref{eq:u} to obtain the physical quantities $\{\psi^i, \bm\tau^i, t_{d_0}^i\}$.

\section{Convergence and performance}
\label{sec:convergence}

\begin{figure}
    \centering
    \includegraphics[width=\linewidth]{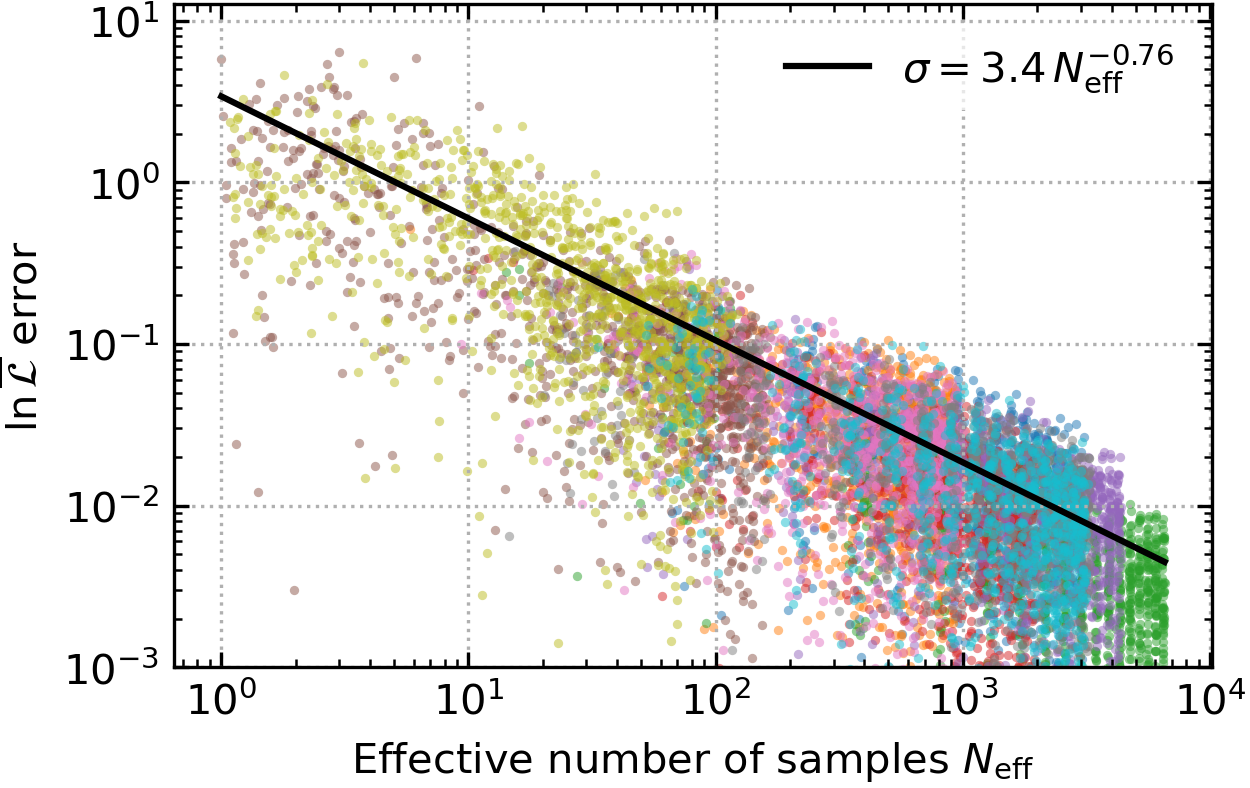}
    \caption{Convergence test of our marginalization algorithm. We show the importance sampling error in the marginalized likelihood $\overline{\mathcal{L}}(d \mid \intrinsic)$ versus effective number of samples in the estimator (Eq.~\eqref{eq:neff}).
    Each point represents a single marginalized likelihood estimate.
    Different colors correspond to different events $d$ and intrinsic-parameter values $\intrinsic$.
    Within each color, points differ in the number of proposal adaptations performed and the realization of extrinsic-parameter importance samples (\S~\ref{sec:proposal}).
    A single fit to the errors is found to describe all examples reasonably well.
    }
    \label{fig:convergence}
\end{figure}

In Fig.~\ref{fig:convergence} we study the accuracy of the marginalization algorithm by comparing multiple estimates to a high resolution result that serves as ground truth.
We find that for different events, intrinsic parameter values, realization of importance samples, and number of optimizations, the effective number of samples remains a good tracer of the error in the computation.
This is important because the effective number of samples can be computed from the available importance weights at negligible cost.
In the figure legend we report the maximum-likelihood fit of a model in which the $\ln\overline{\mathcal{L}}$ errors are normally distributed with a variance that follows a power-law on the effective number of samples.
Notably, very precise estimates of the marginal likelihood can be obtained as needed by increasing the number of samples.

\begin{figure}
    \centering
    \includegraphics[width=\linewidth]{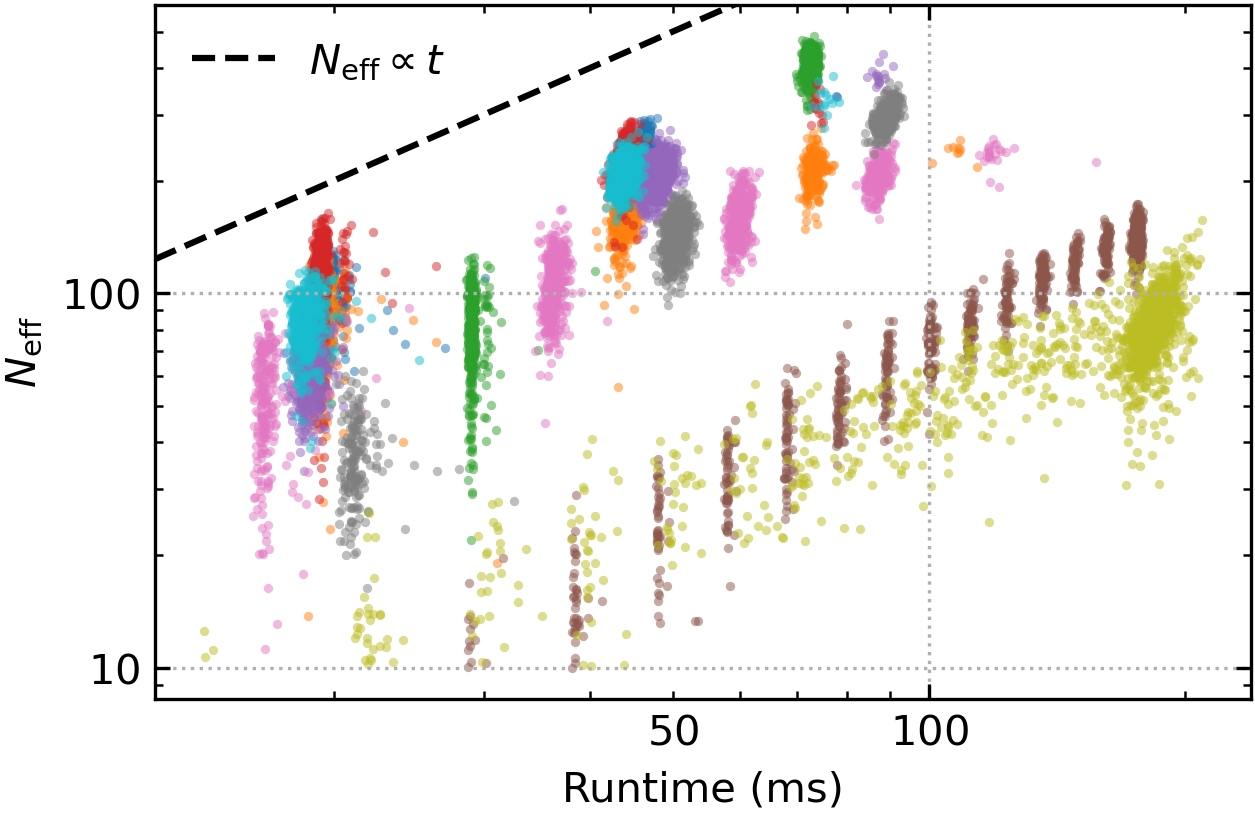}
    \caption{Computational cost of the marginalization over extrinsic parameters.
    We plot the effective number of samples in the importance sampling estimator as a function of the time spent by one CPU core.
    As in Fig.~\ref{fig:convergence}, different colors correspond to different events $d$ and intrinsic-parameter values $\intrinsic$.
    While the computational cost depends considerably on the particular event and parameter values, in most cases $N_{\rm eff} \gtrsim 100$ ($\lesssim 10\%$ uncertainty in $\overline{\mathcal L}$) is achieved within \SI{50}{\milli\second}.
    }
    \label{fig:computation}
\end{figure}

In Fig.~\ref{fig:computation} we show the computational cost of the extrinsic-parameter marginalization using our implementation of the algorithm.
The number of effective samples increases linearly or faster with the computational effort; the latter situation is indicative of cases where the proposal adaptation makes an impact.
We typically achieve 10\% precision within \SI{50}{\milli\second}---with a sizable variance contingent on the event and intrinsic parameter values.

In both Fig.~\ref{fig:convergence} and \ref{fig:computation}, the intrinsic-parameter evaluation points were chosen at random from the chain of proposals made by the \texttt{nautilus} sampler \cite{Lange2023} in a parameter estimation run on data with a synthetic signal injected (see \S\ref{sec:pe} for additional details), to ensure that they are representative of real-world applications.

\section{Applications}
\label{sec:applications}

\subsection{Search for binary mergers}
\label{sec:search}

As argued in \S\ref{sec:introduction}, optimally searching for gravitational wave events requires ranking candidates by their likelihood marginalized over the nuisance parameters of the signal model.
The method introduced in \S\ref{sec:method} allows to marginalize over the extrinsic parameters accounting for higher-order modes, enabling searches sensitive to this type of signals.
Indeed, we have used this algorithm in a search pipeline that incorporates higher modes \cite{Wadekar2023}.

In terms of sensitivity, one could compare this method to a common alternative to marginalization, which is to maximize the likelihood over nuisance parameters.
That approach is reasonably close to optimal in the case where there are a few well-measured parameters.
However, higher-order modes introduce nuanced details in the waveform shape that greatly increase the diversity of waveforms, while at the same time they are a subdominant perturbation for most parameter values.
The diversity of waveforms incurs a large trials factor, to the point where the inclusion of higher modes may in fact degrade the sensitivity of a maximum-likelihood search \cite{Capano2014}.
This degradation happens because the likelihood may be maximized for fine-tuned configurations that are not representative of the generic solutions, whereas marginalization correctly penalizes such configurations.
For example, oftentimes the likelihood is maximized for a nearly edge-on inclination, but this solution is penalized when we marginalize over distance, as highly inclined systems are observable in a smaller volume.
From first principles, with the marginalization statistic the more accurate model (including higher harmonics) is guaranteed to have a better sensitivity by the Neyman--Pearson lemma.

In terms of computational cost, the most expensive inputs to the algorithm are the matched-filtering time series $\timeseries$ for each of the harmonic modes.
For a search, these are computed by means of Fast Fourier Transforms.
This requires constructing the template bank in terms of the individual harmonics, which turns out to be convenient since (in line with the above discussion) it leads to banks of a  $\sim 100$ times smaller size compared to banks of fully specified templates \cite{Wadekar2023tb}.
The covariances $\covariance$ are time-independent (except for slow variations due to the non-stationarity of the noise) and therefore inexpensive.
The efficiency of our marginalization routine enabled us to use it as component of the ranking score on $\sim \num{e7}$ foreground and background (i.e., with artificial time shifts between detectors applied \cite{Usman2016, Venumadhav2019}) triggers in a recent search including higher-order modes \cite{Wadekar2023}, without it becoming a computational bottleneck.

While the likelihood ratio in this work was derived under the assumption of Gaussian noise, a later stage in the pipeline applies a correction for the fact that the empirical noise distribution is not Gaussian \cite{Wadekar2024}.

\subsection{Parameter estimation}
\label{sec:pe}

In this section we demonstrate the applicability of the extrinsic parameter marginalization to parameter estimation.
We use a general purpose stochastic sampler to explore the intrinsic-parameter posterior
\begin{equation}\label{eq:intrinsic_posterior}
    p(\intrinsic \mid d) = \pi(\intrinsic) \overline{\mathcal{L}}(\intrinsic).
\end{equation}
To reconstruct the full distribution, for each intrinsic-parameter sample, we select extrinsic parameters from the conditional posterior $p(\extrinsic \mid \intrinsic, d)$ according to the weights $w_i$ in Eq.~\eqref{eq:importance_marg_like_2}.
The concept and motivation are the same as in \citet{Islam2022}: sampling Eq.~\eqref{eq:intrinsic_posterior} is a lower dimensional problem than the full posterior $p(\theta \mid d)$, therefore will typically take less model evaluations to converge robustly.
The main difference is that here we include higher modes and precession.
Another parameter estimation framework that is based on the marginal likelihood is \texttt{RIFT} \cite{Pankow2015, Lange2018, Wysocki2019, Rose2022, Wofford2023}.
\texttt{RIFT} evaluates the marginal likelihood in parallel on a grid over intrinsic parameters, and then constructs a fast interpolator with which it explores the posterior.
Our algorithm instead runs on a single core and freshly computes the marginal likelihood at every call.
This is rendered possible by the efficiency of our implementation, which computes a marginalized likelihood in $\sim \SI{50}{\milli\second}$.
In comparison, \texttt{RIFT} takes tens of seconds on a GPU or minutes on a CPU \cite{Wysocki2019}.

We perform two tests of this method.
In \S\ref{sec:marginalized_vs_not} we compare it to the more standard strategy of running the sampler on the full parameter space, confirming that we achieve a consistent result on an individual event at a reduced computational cost.
In \S\ref{sec:pp} we generate a large set of synthetic events, and test with P--P plots  that the method recovers the injected parameters consistently across parameter space.

We perform the parameter estimation runs using the \texttt{cogwheel} code \cite{Roulet2022, Roulet2024_cogwheel_code}.
We use the \texttt{IMRPhenomXODE} waveform model, which accounts for precession and the $\{(2, 2), (2, 1), (3, 3), (3, 2), (4, 4)\}$ harmonic modes~\cite{Yu2023}.
The matched-filtering time series $\timeseries$ and covariances $\covariance$ are computed using the heterodyne/relative-binning method \cite{Cornish2013,Zackay2018,Leslie2021}, with the implementation of \cite{Roulet2024}.
We ``fold'' the posterior for the inclination $\theta_{JN}$ as described in \cite{Roulet2022} in order to handle its multimodality (the rest of the parameters identified there as suitable for folding are extrinsic, so the marginalization obviates folding those).
We use the stochastic sampler \texttt{nautilus} \cite{Lange2023} with 1000 live points.
Table~\ref{tab:parameter_values} reports the configuration of the marginalization algorithm we used.

\begin{table}
\caption{Configuration of the algorithm used in \S\ref{sec:pe}.}
\label{tab:parameter_values}
\begin{tabular}{lll}
\toprule
 & Parameter & Value \\
\midrule
$\Delta$ & Time resolution of the mapping  & $2^{-13}$\,s \\
$N_j$ & Number of samples per partial proposal  & 2048 \\
$N^{\rm min}_{\rm eff}$ & Minimum effective number of samples  & 50 \\
$j_{\rm max}$ & Maximum number of proposal adaptations  & 16 \\
$N_\phi$ & Number of phase quadrature points & 128 \\
 & Time series interval around trigger & $\pm \SI{70}{\milli\second}$ \\
\bottomrule
\end{tabular}
\end{table}

\subsubsection{Comparison to no marginalization}
\label{sec:marginalized_vs_not}

As a first sanity check, we infer the parameters of GW190814 \cite{Abbott2020_GW190814}---an event that displays higher modes---in two ways: with our extrinsic marginalization method, or without marginalizing the likelihood and letting the sampler explore the 15-dimensional parameter space.
For the non-marginalized case, we fold the $(\theta_{JN}, \hat\phi_{\rm net}, \phiref, \psi)$ parameters in order to improve the inference efficiency and robustness \cite{Roulet2022}.

\begin{figure}
    \centering
    \includegraphics[width=\linewidth]{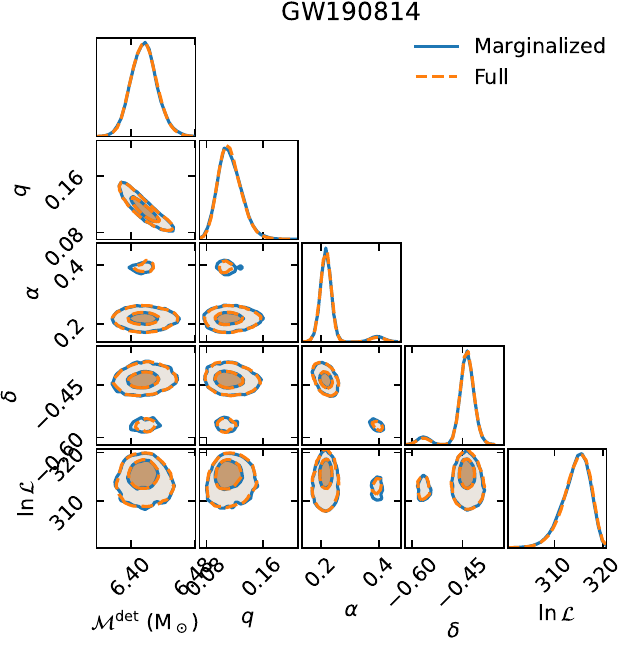}
    \caption{
    Application of our method to parameter estimation.
    In solid blue, we let the stochastic sampler explore the intrinsic parameters following the extrinsic-marginalized posterior, and reconstruct the extrinsic in postprocessing.
    In dashed orange, we explore all 15 intrinsic and extrinsic parameters with the sampler.
    Both achieve similar results, but the marginalized case was $3.5\times$ faster.}
    \label{fig:comparison_to_no_marginalization}
\end{figure}

In Fig.~\ref{fig:comparison_to_no_marginalization} we see that the two methods are in excellent agreement on intrinsic and extrinsic parameters, and the likelihood.
The run with marginalization took \SI{1.6}{h} on one CPU core, while the run without marginalization required \SI{5.5}{h} ($3.5\times$ the cost).
Other events gave similar results: using the marginalization was consistently $\sim3$--5 times faster. 

\subsubsection{Performance on synthetic events}
\label{sec:pp}

\begin{figure*}
    \centering
    \includegraphics[width=\linewidth]{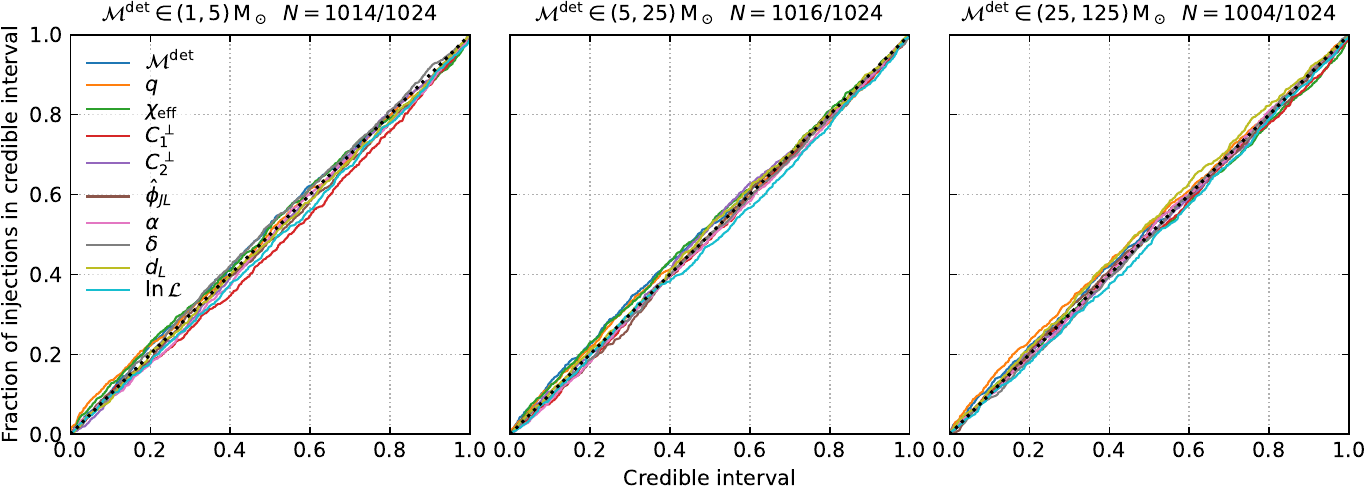}
    \caption{Probability--probability plots for parameter inference on low-mass, medium-mass and high-mass injections.
    The empirical distributions of the percentiles are observed to be uniform (their cumulatives follow a diagonal line), indicating satisfactory performance. Each line corresponds to a different source parameter (the parameters are defined in \cite{Roulet2022}).
    Titles report the fraction of runs that succeeded on the first attempt---the remainder timed out or failed the $\langle h \mid h \rangle > 70$ cut, see text.
    Our method achieves excellent recovery of injection parameters over the wide parameter space that we tested.}
    \label{fig:pp}
\end{figure*}

We further assess the performance of the method by means of probability--probability (P--P) plots, shown in Fig.~\ref{fig:pp}.
That is, we perform a set of injections on Gaussian noise, with source parameters drawn from a prior distribution.
We obtain posterior samples for each injection using the same prior, and test the uniformity of the percentiles $P_\theta$ estimated from the posterior samples for various source parameters $\theta$.
The percentiles represent the probability that a parameter lies below the injected value; in a well-calibrated inference they should follow a uniform distribution over the set of injections:
\begin{equation}
    P_\theta
    \equiv \int_{-\infty}^{\theta_{\rm inj}} \rmd \theta \, p(\theta \mid d) \sim \rm U (0, 1).
\end{equation}
Deviations from uniformity may indicate biases or inaccuracies in our method's performance.

To have more granular information, we partition the parameter space in three bins by detector-frame chirp mass: $\mchirp / \rm M_\odot \in (1, 5)$, $(5, 25)$ or $(25, 125)$. Within each bin, we use a mass prior uniform in detector-frame component masses with a cut in mass ratio $q > 1/20$, a ``volumetric'' spin prior (i.e., isotropic and with $\pi(\chi) \propto \chi^2$ for either dimensionless spin magnitude $\chi$), and uniform in luminosity volume up to $D_{\rm max} = \SI{1.5}{Gpc}$ in the low-mass bin and \SI{15}{Gpc} in the other two.
To have a sample of events more representative of the set of detections, we further impose a cut $\langle h \mid h \rangle > 70$ on the injections.
We do not use this cut during parameter estimation, but reject samples that do not satisfy it in postprocessing.
We use a Hanford--Livingston--Virgo network with average sensitivities from the third observing run.
The parameters are specified at a reference frequency of $\SI{50}{\hertz}$.

We show the results in Fig.~\ref{fig:pp}.
1--2\% of the runs failed (timed out at \SI{24}{\hour} or produced posteriors below the $\langle h \mid h \rangle = 70$ cut we imposed on the injection prior), we exclude these from the plot.
The fraction of runs that succeeded at the first attempt is reported in the figure titles.
We recover satisfactory P--P plots in all parameters and chirp-mass ranges, providing evidence that the posteriors are well calibrated across the parameter space.


The computational cost of the inferences is summarized in Table~\ref{tab:timing}.
The average inference runtimes to produce $\sim \num{e4}$ effective posterior samples range from \SI{0.78}{\hour} (high mass signals) to \SI{2.6}{\hour} (low mass) using a single CPU core.
Our algorithm is thus reasonably efficient and competitive with other state-of-the-art codes.
We find that lower-mass systems take a longer time to run. The reason is threefold: for these systems each waveform generation takes longer, so does the likelihood marginalization, and more likelihood evaluations are needed overall for the sampler to converge.
\texttt{IMRPhenomXODE} is slower for lighter systems because this approximant solves a differential equation for the spin dynamics, and these undergo more precession cycles from a given starting frequency if the masses are smaller.\footnote{The reason is not simply that the waveforms are longer: this would not occur with analytic approximants such as others in the \texttt{IMRPhenom} family, since the evaluation frequencies are independent of waveform duration in the relative binning algorithm \cite{Zackay2018}.}
One likely explanation for the marginalization being less efficient is that low-mass templates have a shorter autocorrelation length, since these systems emit up to higher frequencies.
This allows to measure the arrival time at the detectors better, reducing the target volume of phase space.
The reason why the sampler requires more likelihood evaluations might be related to the prominence of various degeneracies in different regions of parameter space.

The average time for each likelihood marginalization (after the inputs in Eqs.~\eqref{eq:timeseries} and \eqref{eq:covariance} had been generated) was in the range 38--\SI{60}{\milli\second} (depending on the mass bracket) with $N_{\rm eff}^{\rm min} = 50$; this is in line with the estimation from Fig.~\ref{fig:computation}.
The marginalization amounted to approximately 70\% of the overall computational cost: unlike the majority of parameter estimation codes, the cost of waveform generation---while not negligible---was not the dominant bottleneck.
This suggests that somewhat more expensive models could be used without significantly affecting performance.

\begin{table}
\caption{Timing statistics for the same set of parameter inference on injections shown in Fig.~\ref{fig:pp}.
For each chirp-mass range, we report the average inference runtime per event on one CPU core,
the average cost $\tau_{\overline{\mathcal{L}}}$ of each call to the marginalization routine
---which dominates that of the waveform model $\tau_{\texttt{XODE}}$---,
the number $N_{\rm calls}$ of likelihood evaluations performed per event
and the average effective sample size achieved (\texttt{nautilus} produces weighted posterior samples).
}
\label{tab:timing}
\begin{tabular}{cS[table-format=3.2]S[table-format=3.2]S[table-format=3.2]cc}
\toprule
$\mathcal{M} / \rm M_\odot$ & {$\langle$Runtime$\rangle / {\rm h}$} & $\langle \tau_{\overline{\mathcal{L}}}\rangle / {\rm ms}$ & $\langle \tau_{\texttt{XODE}}\rangle / {\rm ms}$ & $\langle N_{\rm calls}\rangle$ & $\langle{\rm ESS}\rangle$ \\
\midrule
$(1, 5)$ & 2.6 & 60 & 23 & \num{9.6e+04} & \num{1.2e+04} \\
$(5, 25)$ & 1.2 & 45 & 8.2 & \num{6.9e+04} & \num{1.2e+04} \\
$(25, 125)$ & 0.78 & 38 & 7.1 & \num{5.2e+04} & \num{9.7e+03} \\
\bottomrule
\end{tabular}
\end{table}


In addition to gauging the consistency of our method, the set of injections and posterior samples we have generated could be utilized for other analyses  as a realistic mock catalog of observations \cite{Fishbach2018, Callister2020, Gray2020, Farah2023}.
With this motivation, we release these data products \cite{Roulet2024b}.

\subsection{Low-latency source localization}
\label{sec:skyloc}

Beyond marginalizing the likelihood, the weighted samples produced by our algorithm can be used to reconstruct the posterior on extrinsic parameters, including the source location.
The computational speed of our marginalization algorithm makes it appealing in the search of short-lived electromagnetic counterparts to gravitational wave signals.
Other algorithms in the literature that can localize a source in low-latency are restricted to quadrupolar waveforms \cite{Singer2016, Pathak2024} or high-mass systems that are unlikely to emit light \cite{Dax2021}.
In contrast, our source localization method works for both high and low mass systems and accounts for higher harmonics.

In this section, we will assume that the spins are aligned with the orbit, which allows us to treat the inclination $\iota$ as an extrinsic (analytic) parameter and infer its value.
We achieve this by including the inclination along with the time, sky location and polarization in the importance sampling (\S\ref{sec:proposal}), and including the full spin-weighted spherical harmonics in the waveform model---i.e., replacing $\rme^{\rmi m \phiref}$ by $_{-2}Y_{\ell m}(\iota, \phiref)$ in Eq.~\eqref{eq:waveform}.

\begin{figure}
    \centering
    \includegraphics[width=\linewidth]{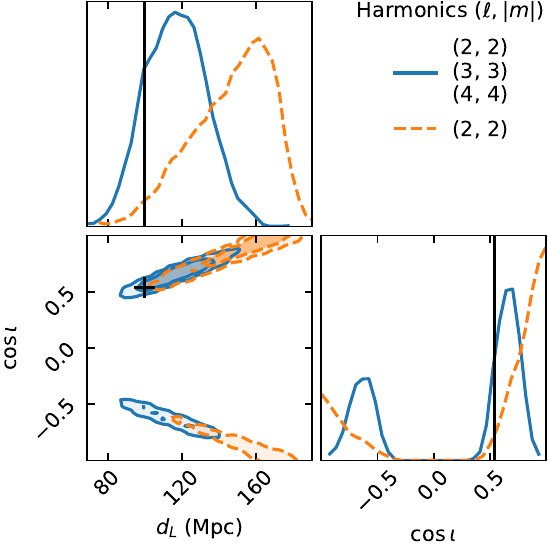}
    
    \vspace{5pt}
    \includegraphics[width=\linewidth]{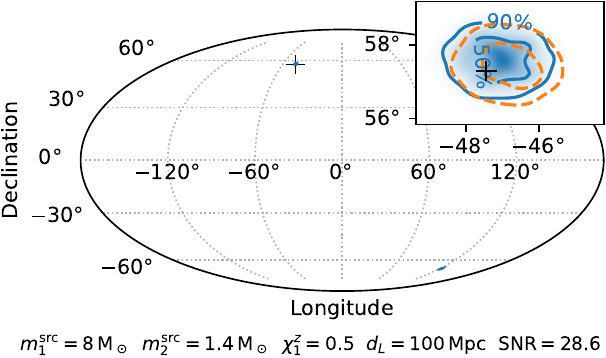}
    \caption{If intrinsic parameters are available from a search pipeline, our method can localize the source in low latency (which is crucial for multimessenger astronomy) while accounting for higher harmonics.
    To illustrate this, we show the source location recovery for a synthetic neutron-star--black-hole signal in a Hanford--Livingston--Virgo network in two cases: using a waveform with higher harmonics (solid blue) or without (dashed orange).
    Injected parameters are indicated with a black cross.
    In both cases, masses and spins are (unrealistically) set to their true values.
    \textit{Top:} Higher modes partially lift the distance--inclination degeneracy, improving the low-latency distance measurement relative to current the state of the art, which only includes the quadrupole.
    \textit{Bottom:} The source is constrained to two disjoint possible arrival directions, the inset zooms in around the correct one.
    }
    \label{fig:skyloc}
\end{figure}

In Fig.~\ref{fig:skyloc} we demonstrate the application of our algorithm to low-latency source localization.
We generate a synthetic signal on Gaussian noise in a Hanford--Livingston--Virgo network with sensitivities typical of the O3 observing run.
We simulate the merger between a \SI{1.4}{M_\odot} neutron star and a spinning \SI{8}{M_\odot} black hole, with dimensionless spin 0.5 aligned with the orbit.
Such system could realistically disrupt the neutron star before merger and produce electromagnetic radiation \cite{Foucart2018}.
We place the source in a sky location with good interferometer response at a distance of \SI{100}{Mpc}, which yields a (recovered) signal-to-noise ratio of 28.6.
We simulate the signal using the \texttt{IMRPhenomXHM} approximant \cite{GarciaQuiros2020} (the aligned-spin limit of \texttt{IMRPhenomXODE}).

We retrieve the extrinsic parameter posterior in two ways: modeling the $(\ell, |m|) = \{(2, 2), (3, 3), (4, 4)\}$ harmonics, or only the $(2, 2)$.
The latter case is intended to represent the current state of the art in low-latency source localization.
At least in this example, we observe that the two results are approximately similar in terms of the sky coordinates.
However, higher harmonics provide a significant help for constraining the source inclination, and thereby the distance.
(A similar phenomenon was reported in the event GW190412 \cite{Abbott2020_GW190412}.)
This hints at the exciting possibility of ruling out some candidate galaxies in the localization region, facilitating the task of identifying a potential counterpart.

Fig.~\ref{fig:skyloc} was produced with 4000 extrinsic-parameter samples, which took \SI{12}{\second} to generate in a single CPU core (after the matched filtering time series and covariances had been generated---those would be provided by a search pipeline).
It would be straightforward to produce multiple smaller batches of samples in parallel if a speedup is desired.

An important caveat in this demonstration is that we treated the intrinsic parameters as known and set them to the true values---in a real application, one would only have access to some noisy estimate from the search pipeline, and would have to marginalize the intrinsic parameters as well.
In particular, intrinsic parameters correlate with the distance, as the loudness of the source depends on the masses and spins.
This further motivates the inclusion of intrinsic parameters in a more detailed reconstruction of the source location.
Higher modes could be of further help in measuring the distance and physical nature of the source, as they also help constrain the intrinsic parameters (by breaking the mass-ratio--effective-spin degeneracy).
Exploring the consequences of this, as well as the extent to which the inclusion of higher harmonics is important in different regions of parameter space, are left to future work.

\section{Conclusions}
\label{sec:conclusions}

We have developed, implemented and tested an efficient algorithm to marginalize the likelihood function of a gravitational wave signal over its extrinsic parameters (and conversely, to sample the posterior conditional on the intrinsic parameters).
The computation assumes Gaussian noise and a quasicircular (non-eccentric) orbit, and works for signals with precession and/or higher-order harmonics.

For precessing signals (spins misaligned with the orbit), we are able to marginalize out six parameters, namely the orbital phase, distance, coalescence time, polarization angle, right ascension and declination. 
For aligned-spin signals, we can additionally marginalize the inclination angle.
We perform the marginalization over distance via a lookup table, over the orbital phase with trapezoid quadrature, and over the remaining extrinsic parameters using adaptive importance sampling.

Our Python implementation of this algorithm typically achieves a $\sim 10\%$ accuracy in \SI{50}{\milli\second} on one CPU core.
We make it available through the software \href{https://github.com/jroulet/cogwheel}{\texttt{cogwheel}}.\footref{footnote:cogwheel_url}

We discussed three applications for this tool: search, parameter estimation, and low-latency localization.

In a search for gravitational wave signals, this algorithm is a key piece in the optimal detection statistic, as it computes the Neyman--Pearson likelihood ratio of the hypothesis that there is a signal with given intrinsic parameters versus Gaussian noise.
This statistic is computed from the time series of data matched-filtered with the individual harmonic modes of the signal (as opposed to the fully specified waveform), which significantly reduces the size of the template bank \cite{Wadekar2023tb}.
It combines data from different detectors coherently, and correctly penalizes fine-tuned configurations.
Our implementation is sufficiently fast that this statistic can be used to rank the large number of foreground and background triggers originating from a search pipeline.

In parameter estimation, by marginalizing the extrinsic parameters we are able to simplify the task of the stochastic sampler: it only needs to explore the intrinsic parameter space, which is lower-dimensional and often has a simpler structure.
We have demonstrated this on thousands of synthetic signals, recovering satisfactory probability--probability plots across the parameter space.
The inference is completed in a one-to-few-hour timescale on a single CPU core, using a waveform model that includes spin-induced precession and higher harmonics.

Finally, we briefly explored the applicability of this algorithm to low-latency source localization, which would be useful in the follow up of electromagnetic counterparts to gravitational wave signals.
For this application we exploit the capability of efficiently sampling the extrinsic-parameter posterior at given intrinsic parameters.
We have shown that accounting for the higher harmonics can make a difference in the recovered distance to the source (by partially lifting the degeneracy with the inclination angle), which suggests the possibility of improving the probability ranking of candidate host galaxies.
Interfacing this routine with low-latency search pipelines and demonstrating its performance on synthetic signals are interesting directions for future work. 

Beyond these applications, this algorithm could be applied to other use cases with relatively straightforward modifications.
For example, the computation of the Bayes factor for a strong-gravitational-lensing hypothesis given multiple candidate images involves a similar integration of the likelihood over the parameters of the signal \cite{Haris2018,Hannuksela2019,Dai2020, Abbott2023_lensing}.

\section*{Acknowledgements}

We thank Ankur Barsode and Srashti Goyal for helpful discussion.

This research has made use of data or software obtained from the Gravitational Wave Open Science Center (\url{gwosc.org}), a service of LIGO Laboratory, the LIGO Scientific Collaboration, the Virgo Collaboration, and KAGRA.
LIGO Laboratory and Advanced LIGO are funded by the United States National Science Foundation (NSF) as well as the Science and Technology Facilities Council (STFC) of the United Kingdom, the Max-Planck-Society (MPS), and the State of Niedersachsen/Germany for support of the construction of Advanced LIGO and construction and operation of the GEO600 detector.
Additional support for Advanced LIGO was provided by the Australian Research Council.
Virgo is funded, through the European Gravitational Observatory (EGO), by the French Centre National de Recherche Scientifique (CNRS), the Italian Istituto Nazionale di Fisica Nucleare (INFN) and the Dutch Nikhef, with contributions by institutions from Belgium, Germany, Greece, Hungary, Ireland, Japan, Monaco, Poland, Portugal, Spain.
The construction and operation of KAGRA are funded by Ministry of Education, Culture, Sports, Science and Technology (MEXT), and Japan Society for the Promotion of Science (JSPS), National Research Foundation (NRF) and Ministry of Science and ICT (MSIT) in Korea, Academia Sinica (AS) and the Ministry of Science and Technology (MoST) in Taiwan.

JR acknowldeges support from the Sherman Fairchild Foundation.
TV acknowledges support from NSF grants 2012086 and 2309360, the Alfred P. Sloan Foundation through grant number FG-2023-20470, the BSF through award number 2022136, and the Hellman Family Faculty Fellowship.
BZ is supported by the ISF, NSF-BSF, a research grant from the Center for New Scientists at the Weizmann Institute of Science and a research grant from the Ruth and Herman Albert Scholarship Program for New Scientists.
MZ is supported by the Canadian Institute for Advanced Research (CIFAR) program on Gravity and the Extreme Universe and the Simons Foundation Modern Inflationary Cosmology initiative.

\appendix

\section{Miscellaneous computational tricks}
\label{app:tricks}

In this technical appendix, we outline steps taken to optimize specific computations, which were omitted from the main text for brevity.

\subsection{Time series relative binning}

During the parameter estimation described in \S\ref{sec:pe}, we generate the matched-filtering time series $\timeseries$ using heterodyne/relative-binning, a method that compresses the data to a few hundred weights $u$ by heterodyning with a reference waveform $h^0$.
Thereafter, computing a matched filter requires the waveform evaluated only on a coarse frequency grid $\{f_b\}$.
Normally, a single reference waveform in each detector (decomposed by modes) would be used.
However, in our application we need to provide $\timeseries$ over a range of $\pm \SI{70}{\milli\second}$, which means that the test waveform can differ from any single reference by hundreds of autocorrelation times.

Following the formulation of \cite{Roulet2024}, we approximate the matched filtering time series as
\begin{align}
    z_{mpdt} &\equiv \int \rmd f
        \frac{\tilde d_d(f) \tilde h_{mp}^*(f)}{S_d(f)}
        \rme^{\rmi 2 \pi f t} \\
    &\approx \sum_{b,m} u_{mtdb} \, \tilde h^*_{mp}(f_b), \\
    u_{mtdb} &= \frac{1}{\tilde h^{0*}_{m}(f_b)} 4 \int \rmd f 
        \frac{\tilde d_d(f) \tilde h_{mp}^{0*}(f)}{S_d(f)}
        \rme^{\rmi 2 \pi f t} s_b(f),
    \label{eq:rb}
\end{align}
where $s_b(f)$ are splines that interpolate the Kronecker delta at the coarse frequency grid:
\begin{equation}
    s_b(f_{b'}) = \delta_{bb'}.
\end{equation}
The key modification we have made is that we have included the time axis in the summary data, which can be interpreted as using multiple reference waveforms, each with a different time shift.
Thus, when generating the time series we are always using an appropriately shifted reference, and we avoid applying a large time shift to the low-resolution waveform.
A similar technique had been used in the precursor work of \cite{Islam2022}.

We note that it is not necessary to use different reference waveforms for different polarizations $p$, as the ratio $\tilde h_{m+}/\tilde h_{m\times}$ is typically a smooth function of frequency.

Previous methods have used the Fast Fourier Transform algorithm to generate the matched filtering time series efficiently \cite{Farr2014}.
However, that would require waveforms sampled at evenly spaced frequencies, which conflicts with the irregular $\{f_b\}$ used in relative binning.

\subsection{Sparse spline representation}

One nuisance associated with having included the time axis in the summary data $u_{mtdb}$ in Eq.~\eqref{eq:rb} is that now the summary is much larger, to the point that it can require a nontrivial amount of computation.

To mitigate this, we accelerate the frequency integrals in Eq.~\eqref{eq:rb} (which, in reality, are matrix multiplications along the fine but discrete frequency axis $f$) by using the B-spline representation of $s_b(f)$.
We arrange the splines into a matrix $S_{bf} \coloneqq s_b(f)$, and express it as
\begin{equation}
    S_{bf} = \sum_{b'} C_{bb'} B_{b'f},
\end{equation}
where $C_{bb'}$ is a square matrix of coefficients and $B_{b'f}$ is a set of B-splines.
The crux is that the $B$ matrix is sparse, which provides a significant speedup.
We compute this decomposition using the \texttt{scipy.interpolate.splrep} implementation \cite{Virtanen2020}.

Other formulations of the heterodyne/relative binning algorithm \cite{Cornish2015, Zackay2018, Cornish2021b, Leslie2021} work with frequency bins, thus, they do not integrate over the full frequency range and never encounter this problem in the first place.
On the other hand, our implementation has the advantage that our approximation of the waveform is smooth over the entire frequency range.

\subsection{Stalling the reference waveform decay}

At the core of the heterodyne/relative binning method is the observation that the ratio $\tilde h(f) / \tilde h^0(f)$ between the test and reference waveforms is a smooth function of frequency.
However, a pathological situation can arise when the merger frequency of the reference waveform is lower than that of the test waveform. Then, the ratio diverges at high frequencies and the computation may become numerically unstable.
We fix this by using a modified reference waveform in which the high frequency part (where the last 1\% of the squared signal-to-noise ratio is accumulated) is set to a nonzero constant, with a smooth cross-fading to prevent artifacts.

\subsection{Waveform time convention}
\label{sec:time_convention}

There is a certain amount of arbitrariness in the convention of what is the ``arrival time'' of a waveform.
The practical importance of this for parameter estimation is that the choice of convention can spuriously correlate the time of arrival and the intrinsic parameters (see e.g. \cite[section V]{Roulet2022}).
Marginalizing over the arrival time, as done in this and other works, in principle makes this problem moot, since the sampler does not need to deal with those correlations.

However, we did find some cases---especially for highly precessing signals---where the time conventions differed so much that the peak of the matched filtering timeseries $\timeseries$ got shifted by more than our $\pm \SI{70}{\milli\second}$ time window as the intrinsic parameters were varied during parameter estimation.
This would cause a complete loss of the signal and bias the inferred parameters.
Even in less extreme scenarios, this shift could produce relative-binning errors if the reference and test waveforms are shifted relative to each other.

To fix this, whenever we generate a waveform we apply a time shift to align it to the relative-binning reference.
We obtain this time shift from a weighted least-squares linear fit to the unwrapped phase difference $\Delta\Phi$ between the two $m=2$ waveforms, as follows.
We estimate the phase difference as
\begin{equation}
    \Delta\Phi(f_b) \coloneqq {\rm unwrap}\left[
        \arg\left(\frac{\tilde h_{2+}(f_b)}{\tilde h^0_2(f_b)}
    \right)\right].
\end{equation}
We take the ratio before the argument to ensure that the (potentially very large) phase accumulated by the waveform largely cancels out with that of the reference, rendering the unwrap possible.
We apply a weighted least squares linear fit to this phase, with inverse variances
\begin{equation}
    \sigma^{-2}_b = \int \rmd f \frac{\big|\tilde h^0(f)\big|\cdot\big|\tilde h_{2+}(f)\big|}{S(f)} s_b(f),
\end{equation}
where we have defined an effective power spectral density through
\begin{equation}
    S^{-1}(f) = \sum_d  S^{-1}_d(f).
\end{equation}
This procedure maximizes the match between the two waveforms over time and phase, under the approximation that $\tilde h_{2+}$ is a small (linear) perturbation of $\tilde h^0_{2}$, and using relative binning to compute the inner product.

We extract the time shift from the slope of the linear fit, and apply it to all the modes and polarizations of the waveform.
This ensures that the peak in the time series will occur near that of the reference waveform, and that relative binning errors are kept to a minimum.
The constant part of the linear fit plays no role in these two problems, so we discard it.
The user is unaffected by this process: in order to facilitate the reconstruction of the signal, we still report the parameter samples in the convention of the original approximant.

\subsection{Memory of previous proposals}

During parameter estimation, the likelihood function is evaluated repeatedly at similar parameter values.
Hence, it is likely that the adapted proposal $P(\bm\tau)$ from one marginalized likelihood call (\S\ref{sec:adaptation}) is also suitable for subsequent calls.
With this heuristic, we aim to accelerate the convergence of the importance sampling integral by averaging the initial proposal in each detector $P_d^{(0)}(\tau)$ (as computed in \S\ref{sec:initial_proposal}) with a ``remembered'' proposal $P_d^{\rm past}(\tau)$.
After each likelihood marginalization call, we update this remembered proposal according to the last iteration of the adaptation:
\begin{equation}
    P_d^{\rm past}(\tau) \leftarrow \frac{P_d^{\rm past}(\tau) + \epsilon P_d^{(J)}(\tau)}{1+\epsilon},
\end{equation}
where $\epsilon$ is a tunable parameter that controls how fast the remembered proposal is updated; we use $\epsilon = \num{e-2}$.

For this procedure to be effective, it is essential that the time convention preserves the time-alignment across waveforms, which we achieve with the method of \S\ref{sec:time_convention}.

\subsection{Pruning phases with low maximum likelihood}
\label{sec:pruning_phases}

The phase quadrature in Eq.~\eqref{eq:phase_quadrature} tries out values of the (very well measured) orbital phase over its full range.
Thus, by construction most of the evaluation points will correspond to waveforms completely inconsistent with the data.
We can save many useless computations of $\overline{\mathcal{L}}_D$ by first computing the (cheaper) quantity
\begin{equation}
    2 \max_D \ln \mathcal L = \frac{\langle d \mid h \rangle^2}{\langle h \mid h \rangle}
\end{equation}
and discarding those values of the orbital phase for which the $\max_D \ln\mathcal L$ falls short of the maximum one by a large amount (we use $\Delta \max_D \ln\mathcal L > 12$).

\subsection{Polarization flip}

Before pruning the phases in \S\ref{sec:pruning_phases}, we can salvage the computation invested in some of the points with the following trick.
The very worst-fitting orbital phases have a large negative $\langle d \mid h \rangle$, indicating that the waveform model is in antiphase with the data.
In that case, we can improve the proposal at negligible cost by applying a shift of $\pi/2$ to the polarization angle whenever $\langle d \mid h \rangle < 0$.
This operation changes the sign of the antenna coefficients \cite{Whelan2013}, and thereby of the strain $h$.
The transformation reads
\begin{equation}
    \begin{pmatrix}
        \psi \\ 
        \langle d \mid h \rangle \\ 
        \langle h \mid h \rangle
    \end{pmatrix}
    \mapsto
    \begin{pmatrix}
        \psi + \pi/2 \\ 
        -\langle d \mid h \rangle \\ 
        \phantom{-}\langle h \mid h \rangle
    \end{pmatrix}.
\end{equation}
This procedure improves $N_{\rm eff}$ to some extent.

\subsection{Foregoing optimization for points with low marginalized likelihood}

Especially during the early phase of parameter estimation, the stochastic sampler explores regions of low likelihood and gradually climbs towards the maximum.
In those regions, samples are either rejected or heavily downweighted, to the point that they are irrelevant for all practical purposes (i.e., the posterior samples and the Bayesian evidence).
While some level of accuracy is desirable, so that the sampler can climb the likelihood surface, for samples with sufficiently low likelihood the target $N^{\rm min}_{\rm eff}$ that we impose on the importance sampling integral can be overly conservative.

To save computations in this case, within each inference run we keep track of the maximum recorded value of $\ln \overline{\mathcal L}(\intrinsic)$ up to that point.
Whenever the estimated $\ln \overline{\mathcal L}$ of the current $\intrinsic$ is lower than the historic maximum by a large value (more than 30) we stop optimizing the proposal even if $N_{\rm eff}$ is low.

\section{Example usage}
\label{app:code}

In this appendix we provide a short snippet of code that illustrates how \texttt{cogwheel} can be used to estimate the parameters of event GW150914 using the algorithm described in this article.

\begin{lstlisting}[language=Python]
import pandas as pd
import matplotlib.pyplot as plt
from cogwheel import data
from cogwheel import gw_plotting
from cogwheel import sampling
from cogwheel.posterior import Posterior

# Directory that will contain parameter
# estimation runs:
parentdir = 'example'

eventname, mchirp_guess = 'GW150914', 30
approximant = 'IMRPhenomXPHM'
prior_class = 'CartesianIntrinsicIASPrior'

# Download data from GWOSC
filenames, detector_names, tgps = \
    data.download_timeseries(eventname)
event_data = data.EventData.from_timeseries(
    filenames, eventname, detector_names, tgps)

# Setup Posterior and Sampler
post = Posterior.from_event(
    event_data, mchirp_guess, approximant,
    prior_class)

sampler = sampling.Nautilus(
    post, run_kwargs=dict(n_live=1000))

rundir = sampler.get_rundir(parentdir)
sampler.run(rundir)  # Will take a while

# Load and plot the samples:
samples = pd.read_feather(
    rundir/sampling.SAMPLES_FILENAME)

gw_plotting.CornerPlot(
    samples, params=sampler.sampled_params,
    tail_probability=1e-4).plot()
plt.savefig(rundir/f'{eventname}.pdf',
            bbox_inches='tight')

\end{lstlisting}

\bibliography{main}

\end{document}